\input phyzzx.tex
\tolerance=1000
\voffset=-0.0cm
\hoffset=0.7cm
\sequentialequations
\def\rl{\rightline}

\def\t1{{\tilde 1}}

\def\t{\theta}

\REF{\GK}{A. Giveon and D. Kutasov, Rev. Mod. Phys. {\bf 71} (1999) 983, [arXiv:hep-th/9802067].}
\REF{\TONG}{D. Tong, [arXiv:hep-th/0509216].}
\REF{\SING}{F. Cachazo, K. Intriligator and C. Vafa, Nucl. Phys. {\bf B603} (2001), [arXiv:hep-th/0103067]; F. Cachazo, S. Katz and C. Vafa, [arXiv:hep-th/0108120]}
\REF{\SUP}{E. Witten, Nucl. Phys. {\bf B507} (1997) 658, [arXiv:hep-th/9706109]; M. Aganagic and C. Vafa, [arXiv:hep-th/0012041].}
\REF{\SHA}{M. Aganagic, C. Beem and S. Kachru, Nucl. Phys. {\bf B796} (2008) 1, arXiv:0709.4277[hep-th].}
\REF{\LAST}{E. Halyo, arXiv:0906.2377[hep-th].}
\REF{\WEI}{E. Weinberg and P. Yi, Phys. Rep. {\bf 438} (2007) 65, [arXiv:hep-th/0609055].}
\REF{\VOR}{F. A. Schaposnik, [arXiv:hep-th/0611028].}
\REF{\DW}{E. R. Abraham and P. K. Townsend, Phys. Lett. {\bf B291} (1992) 85; Phys. Lett. {\bf B295} (1992) 225.}
\REF{\ACH}{A. Achucarro and J. Urrestilla, JHEP {\bf 0408} (2008) 050, [arXiv:hep-th/0407193].}
\REF{\KAL}{G. Dvali, R. Kallosh and A. Van Proeyen, JHEP {\bf 0401} (2004) 035, [arXiv:hep-th/0312005].}
\REF{\COL}{A. Collinucci, P. Smyth and A. Van Proeyen, JHEP {\bf 0702} (2007) 060, [arXiv:hep-th/0611111].}
\REF{\COS}{E. Halyo, JHEP {\bf 0403} (2004) 047, [arXiv:hep-th/0312268].}
\REF{\DOUG}{M. Douglas, [arXiv:hep-th/9512077]; J. Geom. Phys. {\bf 28} (1998) 255, [arXiv:hep-th/9604198].}
\REF{\THO}{G. 't Hooft, Nucl. Phys. {\bf B79} (1974) 276.}
\REF{\POL}{A. M. Polyakov, JETP Lett. {\bf 20} (1974) 194.}
\REF{\DOU}{M. Douglas, JHEP {\bf 9707} (1997) 004, [arXiv:hep-th/9612126].}
\REF{\ABR}{A. A. Abrikosov, Sov. Phys JETP {\bf 5} (1957) 1174.}
\REF{\NIE}{H. B. Nielsen and P. Olesen, Nucl. Phys. {\bf B61} (1973) 45.}
\REF{\KIB}{M. B. Hindmarsh and T. W. B. Kibble, Rep. Prog. Phys. {\bf 58} (1995) 477, [arXiv:hep-ph/9411342].}
\REF{\CST}{E. Halyo, arXiv:0906.2587[hep-th].}
\REF{\NAB}{S. Mandelstam, Phys. Lett. {\bf B53} (1975) 476; H. J. de Vega, Phys. Rev. {\bf D18} (1978) 2932.}
\REF{\ETO}{M. Eto et. al., Phys. Rev. Lett. {\bf 96} (2006) 161601, [arXiv:hep-th/0511086]; Phys. Rev. {\bf D74} (2006) 065021; [arXiv:hep-th/0607070].}
\REF{\SEMI}{A. Achucarro, and T. Vachaspati, Phys. Rev. {\bf D44} (1991) 3067; Phys. Rev. {\bf D44} (1991) 3067; Phys. Rept. {\bf 327} (2000) 347, [arXiv:hep-ph/9904229].}
\REF{\AUZ}{R. Auzzi et. al., Nucl. Phys. {\bf B813} (2009) 484; arXiv:0810.5679[hep-th].}
\REF{\ONE}{M, Eto et. al., Phys. Rev. Lett. {\bf 93} (2004) 161601, [arXiv:hep-th/0404198].}
\REF{\TWO}{Y. Isozumi, M. Nitta, K. Ohashi and N. Sakai, Phys. Rev. {\bf D70} (2004) 125014, [arXiv:hep-th/0405194].}
\REF{\HAN}{A. Hanany and E. Witten, Nucl. Phys.{\bf B492} (1997) 152, [arXiv:hep-th/9611230].}
\REF{\DIA}{D. E. Diaconescu, Nucl. Phys. {\bf B503} (1997) 220, [arXiv:hep-th/9608163].}
\REF{\AMI}{A. Hanany and D. Tong, JHEP {\bf 0404} (2004) 066, [arXiv:hep-th/0403158]; Comm. Math. Phys. {\bf 266} (2006) 647, [arXiv:hep-th/0507140].}
\REF{\THR}{M. Eto et. al., Phys. Rev. {\bf D71} (2005) 125006, [arXiv:hep-th/0412024].}
\REF{\MOD}{M. Eto et. al., Phys. Rev. {\bf D76} (2007) 105002, arXiv:0704.2218[hep-th].}


\singlespace
\rl{SU-ITP-09/30}
\pagenumber=0
\normalspace
\medskip
\bigskip
\titlestyle{\bf{Solitons on Singularities}}
\smallskip
\author{ Edi Halyo{\footnote*{e--mail address: halyo@stanford.edu}}}
\smallskip
\centerline {Department of Physics} 
\centerline{Stanford University} 
\centerline {Stanford, CA 94305}
\smallskip
\vskip 2 cm
\titlestyle{\bf ABSTRACT}

We describe solitons that live on the world--volumes of D5 branes wrapped on deformed $A_n$ singularities fibered over $C(x)$. 
We show that monopoles are D3 branes wrapped on an $A_1$ singularity and stretched along $C(x)$. F and D--term strings are D3 branes wrapped on a node of an $A_2$ singularity that is deformed and resolved respectively. Domain walls are D5 branes wrapped on a deformed $A_3$ singularity and stretched along $C(x)$.

\singlespace
\vskip 0.5cm
\endpage
\normalspace

\centerline{\bf 1. Introduction}
\medskip

In field theory, solitons are important objects because they teach us about the semi--classical and nonperturbative physics. The same is true for supersymmetric field theories that live on the
world--volumes of different D--branes. Since, most if not all field theories can be obtained by some configuration of D--branes[\GK], we expect to find solitons in many D--brane
configurations. In fact, a large body of work already exists on solitons in supersymmetric field theories that can be obtained in intersecting brane models[\TONG]. 

An alternative way to examine world--volume theories of D--branes is to locate them on singularities[\SING]. For example, one can wrap D5 branes on (the nodes of) $A_n$ singularities and examine
the world--volume theories in the $3+1$ noncompact directions. The resulting world--volume gauge groups and couplings, matter contents and superpotentials 
are well--understood[\SING,\SUP]. By deforming and resolving the different nodes of the singularity, and taking large volume limits, one can obtain many different field theories. 
It has been shown that some of these lead to supersymmetry breaking[\SHA,\LAST].

In this paper, we describe solitons of different dimensions that live on $A_n$ singularities. In particular, we concentrate on monopoles[\WEI], vortices[\VOR] and domain walls[\DW] that can live on the 
(noncompact) world--volumes of D5 branes wrapped on nodes of deformed $A_n$ singularities fibered on $C(x)$. We obtain the different soliton solutions by either deforming or resolving the nodes of the singularity. 

For example, monopoles can be obtained by a wrapping D5 branes on the simplest singularity, i.e. an $A_1$ with a single node that is neither deformed nor fibered. Monopoles arise when the D5 branes are separated along the transverse $C(x)$ which corresponds to the Coulomb branch of the world--volume theory. In the brane setup, 
monopoles are described by D3 branes that are wrapped on the node and stretched along $C(x)$ between any two D5 branes. Vortices arise from wrapping D5 branes on an $A_2$ singularity with two nodes since they require charged fields that spontaneously break the $U(1)$ gauge group and these arise from strings stretched between branes on the two separate nodes. In order to realize the spontaneous symmetry breaking of the $U(1)$, the $A_2$ has to be deformed and fibered over $C(x)$.
In the brane setup, vortices correspond to D3 branes that wrap a node but stretch along one of the the noncompact D5 world--volume directions. We show that these correspond to F--term strings[\ACH]. D--term strings[\KAL,\COL,\COS] are obtained by a resolution[\DOUG] of the node in addition to its deformation. This leads to an anomalous D--term on the world--volume theory which gives rise to D--term strings.
Non--Abelian vortices arise if there are multiple D5 branes wrapped on the different nodes of the singularity. Domain walls, on the other hand, require at least three nodes and therefore the smallest singularity that gives rise to them is a deformed and fibered $A_3$ singularity. The deformations 
lead to isolated vacua and therefore to domain walls. If, in addition, the node is resolved, there are semi--local vortices[\ACH] connected to the domain walls. We show that these domain walls are D5 branes wrapped on a node and stretched along $C(x)$.
Our treatment is purely semi--classical and we are mainly interested in the existence of different solitons that live on singularities. In particular, in this paper, we are not concerned about interactions between solitons, their moduli spaces, world--volume theories or other quantum properties. 

The paper is organized as follows. In section 2, we describe the monopole solution that lives on an $A_1$ singularity. We obtain monopoles on the world--volume theory
and show that their properties match those of wrapped D3 branes. In section 3, we describe F and D--term strings which are vortex solutions in the world--volume theory. We show that
these correspond to D3 branes wrapped on a deformed (resolved) and fibered $A_2$ singularities and stretched along one world--volume direction. We also show that wrapping multiple D5 branes on the singularity leads to non--Abelian vortices. Section 4 contains the 
description of domain walls as solutions to the world--volume theory with isolated vacua and as D5 branes wrapped on a deformed and fibered $A_3$. Section 5 contains a discussion of our results and our conclusions.

\bigskip
\centerline{\bf 2. Monopoles}
\medskip

We begin with the description of monopoles[\WEI,\THO,\POL] which are localized solitons in the $3+1$ dimensional theory. Monopoles can be obtained on the
simplest singularity, namely an $A_1$ (which is neither deformed nor fibered over the transverse complex space $C(x)$). Consider the $A_1$ singularity given by
$$uv=(z-z_0)(z-z_0) \eqno(1)$$
where $z_0$ is a constant and the singularity which consists of one node (or $S^2$) is located at the origin of $C(x)$. If we wrap $N$ D5 branes on this singularity,
the world--volume theory on the noncompact $3+1$ dimensions becomes an ${\cal N}=2$ supersymmetric $U(N)$ gauge theory. In ${\cal N}=1$ terms, the theory has
a $U(N)$ gauge superfield together with a hypermultiplet in the adjoint representation, $\phi$. The gauge coupling is given by


$${{4 \pi} \over g^2}={V \over {(2 \pi)^2 g_s \ell_s^2}} \qquad  \eqno(2)$$
where $g_s$ and $\ell_s$ are the string coupling and length respectively. $V$ is the ``stringy volume''[\DOU] given by $V=(2 \pi)^4 \ell_s^4(B^2+r^2+\alpha^2)^{1/2}$ with
$$B=\int_{S^2} B^{NS}   \qquad  r^2=\int_{S^2} J \eqno(3)$$
i.e. $B$ is the NS-NS flux through the node and $r^2$ is the volume of the blown-up $S^2$. The deformations of the singularity are parametrized by 
$\alpha$ which are related to the F--term for $\phi$ in field theory. There may be a superpotential for $\phi$ 
determined by the deformation data [\SUP] 
$$W(\phi)=\int^{\phi} (z_1(x)-z_2(x)) dx \eqno(4)$$
where $z_{1,2}$ are the zeros of the different factors in eq. (1). Clearly, for our simple $A_1$ singularity $z_1=z_2=z_0$ and therefore there is no 
superpotential for $\phi$.

Since $\phi$ has a (classically) flat potential, the VEV of $\phi$ is completely free and we may write it as
$$<\phi>=diag(a_1,\ldots,a_N) \eqno(5)$$
where in general all $a_i$ are different. This corresponds to the Coulomb branch of the world--volume theory in which (excluding the center of mass $U(1)$)
the gauge group is broken $SU(N) \to U(1)^{N-1}$. At a generic point in the Coulomb branch, the world--volume theory consist of $(N-1)$ massless photons and scalars
in addition to ${1 \over 2}N(N-1)$ massive gauge bosons with mass $M^2=(a_i-a_j)^2$ and their supersymmetric partners. Since the vacua described by the VEVs
in eq. (5) are points in the space $SU(N)/U(1)^{N-1}$, they are classified by the maps $S^2_{\infty} \to SU(N)/U(1)^{N-1}$ which fall into different topological
sectors characterized by
$$\Pi_2(SU(N)/U(1)^{N-1})\simeq\Pi_1(U(1)^{N-1})\simeq {\bf Z}^{N-1} \eqno(6)$$
As a result, when we wrap $N$ D5 branes on the simple $A_1$ singularity we obtain $N-1$ types of monopoles each charged under one of the $N-1$ unbroken $U(1)$s 
in the Coulomb branch.

For the simplest example of a monopole consider wrapping two D5 branes on the $A_1$ singularity given by eq. (1). Then, at a generic point in the Coulomb branch given by $<\phi>=(0,a)$,
$SU(2) \to U(1)$. Since $SU(2)/U(1) \simeq S^2$ and $\Pi_2(S^2) \simeq {\bf Z}$ we obtain the t' Hooft--Polyakov monopole.
In the world--volume field theory, the monopole is described by 
$$\phi={{{\hat r_i} \sigma_i} \over r}(ar coth(ar)-1) \qquad A_{\mu}=-\epsilon_{ijk}{{{\hat r_i} \sigma_i} \over r} \left(1-{{ar} \over {sinh(ar)}} \right) \eqno(7)$$
with magnetic charge, $g_m$, and field
$$g_2 g_m=2 \pi n \qquad B_i={g_m \over {4 \pi r^2}}{\hat r_i} \eqno(8)$$
where $n$ is the topological (magnetic) charge of the monopole.
The mass of a monopole (of charge $n$) is 
$$m_m={{4 \pi} \over g_2^2} an \eqno(9)$$
This monopole is actually a D3 brane wrapped on the second node with volume $V_2$ and stretched between the two wrapped D5 branes at $x=0$ and $x=a$. The mass of such a brane is (with $d=2 \pi \ell_s^2 a$)
$$m=T_{D3} V_2 d={V_2 \over {(2 \pi)^3 g_s \ell_s^4}} (2 \pi a \ell_s^2)={{4\pi} \over g_2^2} a \eqno(10)$$
which exactly matches the monopole mass in eq. (9) for $n=1$. We see that the topological (or magnetic) charge of the monopole corresponds to the winding number of the D3 brane on $S_2^2$.   

This model also contains dyons that carry both electric and magnetic charges. In the above field theory, W bosons have mass $m_W=a$. In fact, W bosons are fundamental strings stretched between the two D5 branes wrapped on the nodes at $x=0$ and $x=a$. 
Thus, a dyon with charge $(p,q)$ 
is a bound state of $p$ fundamental strings and $q$ wrapped D3 branes stretching from $x=0$ to $x=a$. This bound state has a mass given by
$$m_d=\sqrt{p^2 a^2+ q^2{{4\pi a} \over g_2^2} } \eqno(11)$$
which is precisely the dyon mass expected in field theory.
It is well--known that a fundamental string can be bound to D3 brane (on a torus). This is simply a configuration T--dual to the bound state between a fundamental string and a D--string. 
However, in our case the D3 brane is not wrapped on a torus but on $S^2$ which does not allow T--duality. Nevertheless, the above configuration is the only candidate with a mass that 
matches the expected dyon mass. It would be interesting to understand this bound state from the wrapped D3 world--volume theory point of view and verify our result. 

In the more general case with $N$ D5 branes wrapped on the $A_1$ singularity, the $N-1$ types of monopoles that arise in the Coulomb branch defined by the VEV
in eq. (5) have masses
$$m_i={{4 \pi} \over g^2} (a_{i+1}-a_i)n \eqno(12)$$

The field theory described by the superpotential in eq. (5) has ${\cal N}=2$ supersymmetry. The monopoles above break half the supersymmetry and therefore are BPS. As a result, there 
are no corrections to the monopole solution and mass in eqs. (7)-(9) since they are protected by supersymmetry. 

We can easily generalize the above results to the cases of more complicated singularities such as $A_n$ which has $n$ nodes ($S^2$s) given by
$$uv=(z-z_0)^n \eqno(13) $$
In this case we can wrap $N_i$ D5 branes on the $i$th node and obtain a $3+1$ dimensional, ${\cal N}=2$ supersymmetric world--volume theory with gauge group 
$\Pi_i U(N_i)$. The gauge couplings $g_i$ are given by eqs. (2) and (3) which are, in principle, different for each node. In addition each node has a 
hypermultiplet $\phi_i$
in the adjiont of $U(N_i)$ with VEVs that parametrize the Coulomb branch. At a generic point in the Coulomb branch, by the above arguments we obtain $N_i-1$
monopoles at each node giving a total of $\Pi_i (N_i-1)$ monopoles. (Here we assume that the bifundamental fields that arise from strings stretched between the different nodes are massive and decouple from the low energy physics which is true at a generic point in the Coulomb branch.)


\bigskip
\centerline{\bf 3. Vortices}
\medskip

In this section, we consider one dimensional solitons, namely vortices in $3+1$ dimensions[\VOR,\ABR,\NIE]. We obtain F and D--term strings and describe them in terms of D3 branes wrapped on nodes of deformed and resolved $A_2$ singularities (fibered over $C(x)$) respectively. We generalize these results to the case of non--Abelian vortices by considering multiple D5 branes wrapped on the nodes of the singularity. In order to obtain vortices a $U(1)$ gauge group must be spontaneously broken
by the VEV of a charged field. Such charged fields (in the bifundamental representation of the gauge groups) only arise from strings stretched between branes wrapped
on two separate nodes of a singularity. Therefore, the smallest singularity that leads to vortices is $A_2$ which has two nodes.

\medskip
{\bf 3.1. F--term Strings:} 
In order to describe vortices that live on singularities, we consider a deformed $A_2$ singularity fibered on $C(x)$ described by
$$uv=(z-mx)(z+mx)(z+m(x-2a)) \eqno(14)$$
Wrapping one D5 brane on each node ($S^2$) of this singularity leads to the gauge group, $U(1)_1 \times U(1)_2$, and the matter content of two 
bifundamentals $Q_{12},Q_{21}$ and two singlets $\phi_{1,2}$ 
with a superpotential given by[\SUP]
$$W=m \phi_1^2-2ma \phi_2 +Q_{12}Q_{21}(\phi_2-\phi_1) \eqno(15)$$
We can decouple $U(1)_1$ by taking the volume of the first node $V_1$ to be very large, i.e. $V_1>>\ell_s^2$.
We note that, now there is an F--term for the massless field $\phi_2$, $F=-2ma$, in the superpotential. $\phi_1$ is massive and decouples at low energies, $E<<m$. It can be 
integrated out by setting its F--term
$$F_{\phi_1}=2m \phi_1-Q_{12}Q_{21} \eqno(16)$$ 
to zero. The low--energy superpotential becomes
$$W=\phi_2(Q_{12}Q_{21}-2ma)-{(Q_{12}Q_{21})^2 \over {4m}} \eqno(17)$$
and gives rise to the F--terms 
$$F_{\phi_2}=Q_{12}Q_{21}-2ma \eqno(18)$$
$$F_{Q_{12}}=\phi_2 Q_{21}-{{Q_{12}Q_{21}^2} \over {2m}} \eqno(19)$$
$$F_{Q_{21}}=\phi_2 Q_{12}-{{Q_{12}^2Q_{21}} \over {2m}} \eqno(20)$$
In addition, there is the D--term for $U(1)_2$
$$D=|Q_{12}|^2-|Q_{21}|^2 \eqno(21)$$
F and D--terms vanish in the supersymmetric vacuum with $|Q_{12}|=|Q_{21}|$ and
$$Q_{12}Q_{21}=2ma \qquad \phi_2=a \eqno(22)$$
In this vacuum, eq. (16) gives $\phi_1=a$ so the singlet VEVs are equal.
We see that $U(1)_2$ is spontaneously broken by the $Q_{12},Q_{21}$ VEVs since they carry charges $1,-1$ respectively. As a result, the photon gets a mass of $2g_2 \sqrt{ma}$ whereas the matter fields have
masses $m_Q=m_{\phi_2}=\sqrt{2ma}$. As usual, topological considerations imply that the spontaneous breaking of $U(1)_2$ leads to vortex solutions.

Far away from the core of the vortex, at large $r$, the solution is 
$$ Q_{12}=Q_{21}^{\dagger}=\sqrt{2ma}e^{i n \theta}  \qquad  A_{\theta}={n \over {gr}} \qquad  F_{\mu \nu}=0 \eqno(23)$$
where $n$ is the topological winding number. The vortex (along the $z$ direction) has a metric which has a conical singularity
$$ds^2=-dt^2+dz^2+dr^2+r^2 \left (1+{{2man} \over M_P^2} \right)d \theta^2 \eqno(24)$$
Near the core of the vortex, at small $r$, the solution is
$$Q_{12}=Q_{21}=0 \qquad  A_{\theta}={M_P^2 \over {2mag}} \left(1-cos \left({{2mag} \over M_P} \right)r \right) \qquad  \eqno(25)$$

It is well--known that the vortex with winding number $n$ carries a magnetic flux of
$$\Phi_n=\int B_z dxdy=2 \pi n \eqno(26)$$
i.e., the winding number is the magnetic flux.  The tension of the vortex is
$$T_n=2 \pi F n =4 \pi ma n \eqno(27)$$
In order for the vortex to exist at low energies, $E<<m$, we need $T<<m^2$ which means we need to assume $a<<m$. The vortex tension arises due to the nonzero F--term in eq (16). Thus, eq. (23)
in fact describes an F--term string[\ACH]. This vortex has a size (width) given by
$$w \sim {1 \over {g_2 \sqrt{F}}} \sim {1 \over {g_2 \sqrt{2ma}}} \sim \sqrt{\pi} \ell_s \eqno(28)$$
where we used eqs. (2) and (30) below. We see that the vortex size is about the string length and independent of the parameters of the field theory. This is due to the identical dependence
of $g_2^{-2}$ and the F--term on $V_2$ and $g_s$.

The vortex we described above is actually a D3 brane wrapped on the second node, $S_2^2$. We can find the relation between the F--term, $F=2ma$, and the ``stringy volume'' of the 
second node $V_2$ by equating the energy of the D5 brane wrapped on $S_2^2$ to the vacuum energy in the field theory for vanishing VEVs,
$${1 \over 2} g_{2}^2 F^2= T_{D5} V_2={V_2 \over {(2 \pi)^5 g_s \ell_s^6}} \eqno(29)$$
The factor of $g_2^2/2$ above is due to a subtlety related to the normalization of $\phi_2$. The normalization that is common in the literature which we used in the above superpotentials 
has a hidden factor of $g_2/\sqrt{2}$ for every factor of $\phi_2$. This is the reason for the absence of any coupling constant in the superpotential, e.g. in eq. (15). In order to find the relation between $F$ and $V_2$ we need 
to restore these factors of $g_2$. We do this only here since the normalization of $\phi_2$ does not affect any of our other results.
Using eq. (2) for $g_2$ we find
$$F=2ma={V_2 \over {(2 \pi)^4 g_s \ell_s^4}} \eqno(30)$$
which establishes the relation between $F$ and $V_2$. Then, the tension of a D3 brane wrapped on $S_2^2$ is 
$$T_F=T_{D3} V_2={V_2 \over {(2 \pi)^3 g_s \ell_s^4}}=2 \pi F =4 \pi ma \eqno(31)$$
which is precisely the tension of the vortex with $n=1$. This shows that the F--term string we found in field theory is a D3 brane wrapped on $S_2^2$ and stretched along one of the 
noncompact world--volume directions. The topological charge $n$ is simply the number of times the D3 brane wraps $S_2^2$. 

A D3 brane inside a D5 brane constitutes a generalization of a magnetic flux tube[\DOU]. This is usually shown using the coupling between the spacetime RR potential that 
couples to the D3 brane and the 
world--volume gauge field strength on the D5 brane. Alternatively, this configuration is T--dual to a D1 brane inside a D3 brane which is known to represent a magnetic flux tube.
We find that after wrapping both branes on $S_2^2$, the wrapped D3 brane carries one unit
of magnetic flux as expected from a vortex string. This is a little surprising since there is no T--duality on $S^2$ and we cannot directly connect our brane configuration to that of a D1 brane
inside a D3 brane. It would be interesting to resolve this problem by examining the world--volume theory of a D5 brane wrapped on an $S^2$.


Note that the superpotential in eq. (15) has ${\cal N}=1$ supersymmetry and the F--term string is not BPS. Therefore, we expect the string solution and tension to recieve corrections.
However, we expect the string to be stable due to conservation of topological charge. As we mentioned above, the ${\cal N}=1$ supersymmetry is a result of fibering $A_2$ over $C(x)$ 
which manifests itself through singlet masses in the superpotential. Consider the deformed but not fibered $A_2$ singularity
$$uv=(z+a)(z+2a)(z+a) \eqno(32)$$
where $a_i$ are $x$ independent and satisfy $\alpha_i=z_{i+1}-z_i$ with $\alpha_i=\int_{S_i^2}B^{NS}$, i.e. they paramatrize the deformation due to NS flux through the nodes. 
(Unlike above, here the parameter $a$ has dimension 2.) This singularity gives rise to the superpotential
$$W=Q_{12}Q_{21}(\phi_2-\phi_1)+a\phi_1-a\phi_2 =(Q_{12}Q_{21}-a)(\phi_2-\phi_1)\eqno(33)$$
Now, we can take $\phi_1 \not = \phi_2$ so that $U(2) \to U(1)_1 \times U(1)_2$. Then supersymmetry requires $Q_{12}Q_{21}=a$. Decoupling $U(1)_1$ as before, we see that
$U(1)_2$ is spontaneously broken and the model has F--term strings with tension $T_F=2 \pi a n$. These are BPS strings since they exist in an ${\cal N}=2$ supersymmetric model. As above,
they are described by D3 branes wrapped on $V_2$. 

The model described by eq. (15) also contains monopoles. This is not surprising since inside the vortex the Abelian group $U(1)_2$ remains unbroken.
In addition, since $Q_{12}=Q_{21}=0$ inside the vortex, the singlets, $\phi_1,\phi_2$ may have different VEVs. We find that near
the core of the vortex, $\phi_1=0$ whereas $\phi_2$ is free. These monopoles have mass $m_m=4 \pi \phi_2 /g_2^4$. Since
$U(1)_2$ is broken outside the vortices and is restored only in their core, these monopoles are not free but confined by the strings which are flux tubes. In this case, the probability for
monopole--anti monopole creation determines whether the vortices are short flux tubes confining monopoles or long cosmic strings[\VOR,\KIB]. The probability is given by
$$P \sim exp(-\pi m_m^2/T_F) \sim exp \left(-{{4 \pi^2} \over {g_2^2}} {\phi_2^2 \over ma}\right) \eqno(34)$$
Clearly, we can make this probability as small as we want by taking $\phi_2>>\sqrt{ma}$. This choice stabilizes the vortices against monopole--anti monopole pair creation and leads to long
cosmic F--strings[\KIB,\CST].

\medskip
{\bf 3.2. D--term Strings:}
We now show that D--term strings[\KAL,\COS] can also live on singularities. Consider another deformed $A_2$ singularity fibered on $C(x)$ given by
$$uv=(z-mx)(z+mx)(z+mx) \eqno(35)$$
The gauge group is again $U(1)_1 \times U(1)_2$ and the matter content consists of two singlets $\phi_1,\phi_2$ and one pair of bifundamentals
$Q_{12},Q_{21}$ with the superpotential
$$W= m\phi_1^2+Q_{12}Q_{21} (\phi_2-\phi_1)  \eqno(36)$$
As before, we decouple $U(1)_1$ by taking the volume of the first node to be very large in string units.
At low energies $E<<m$, $\phi_1$ decouples (with vanishing VEV) and we are left with
$$W= \phi_2 Q_{12}Q_{21} \eqno(37)$$
In addition, we blow up the second node which gives rise to an anomalous D--term[\DOUG]
$$\xi=\int_{S_2^2} J \eqno(38)$$
where $J$ is the Kahler form on $S^2$. This blow--up of the second node is the main difference between D--term strings and F--term strings described in the previous section.
The D--term for $U(1)_2$ becomes
$$D_2=|Q_{12}|^2-|Q_{21}|^2+ \xi \eqno(39)$$
We see that a supersymmetric vacuum now requires at least a nonzero VEV for $Q_{21}$ which breaks $U(1)_2$ spontaneously, i.e $|Q_{21}|^2=\xi$ and $|Q_{12}|=\phi_2=0$. 
As before, the spontaneous breaking of $U(1)_2$ means that the theory has vortex solutions. These can be obtained
from eqs. (23)-(27) by replacing $F=2ma$ with $\xi$. These vortices are D--term strings that carry $n$ units of magnetic flux and have tension $T_D= 2 \pi \xi n$.

Like the F--term string, the D--term string is also a D3 brane wrapped on the second node. The argument is identical to the one we gave above for F--term strings with the replacement of 
$F=2ma$ by $\xi$. As a result, we find the relation
$$\xi={V_2 \over {(2 \pi)^4 g_s \ell_s^4}} \eqno(40)$$
between the anomalous D--term and the blow--up volume $V_2$. The tension of a D3 brane wrapped on $V_2$ is
$$T_D=T_{D3} V_2={V_2 \over {(2 \pi)^3 g_s \ell_s^4}}=2 \pi \xi\eqno(41)$$
which matches that of the D--term string with $n=1$. For these vortices to exist at low energies, $E<<m$, we need to assume $\sqrt{\xi}<<m$.

D--term strings are BPS even in ${\cal N}=1$ supersymmetric models like the one described by eq. (36)[\ACH,\KAL]. Therefore, their solution and tension do not receive any corrections. 
We can obtain D--term strings in ${\cal N}=2$ supersymmetric models by wrapping the D3 branes on untwisted $A_2$ singularities such as the one given by eq. (31) with a blown up second node. 

It is easy to show that there are monopoles living inside D--term strings; in other words D--term strings confine monopoles. The argument is identical to the one for F--term strings.
Inside a D--term string, $Q_{12}=Q_{21}=0$ and therefore we can have a nonzero $\phi_2$; in fact $\phi_2$ is free. Since near the core, $\phi_1=0$, any nozero $\phi_2$ gives rise to monopoles 
with mass $m_m= 4 \pi \phi_2/g_2^2$. Modifying eq. (34) for the probability for monopole--anti monopole creation, we find that the stability of D--term strings can be guaranteed by taking $\phi_2>>\sqrt{\xi}$.

\medskip
{\bf 3.3. Non--Abelian Vortices:}
Non--Abelian vortices[\NAB,\ETO] can also be obtained by wrapping multiple D5 branes on the nodes of the deformed $A_2$ singularity. 
In order to get a non--Abelian vortex we can wrap $N_f$ and $N_c$ D5 branes on the first and second nodes respectively resulting in a $U(N_f)\times U(N_c)$ gauge group.
Then the bifundamentals $Q_{12}$, and $Q_{21}$ are in the $(N_f, {\bar N_c})$ and $({\bar N_f}, N_c)$ representations of the gauge group respectively. 
If we take the volume of the first node, $V_1$, to be very large, i.e. $V_1>>\ell_s^2$ the $U(N_f)$ coupling given by eq. (2)
becomes very small. As a result, the non--Abelian gauge dynamics decouples and $U(N_f)$ becomes a global symmetry. Then, $Q_{12},Q_{21}$ become $N_f$ flavors in the $N_c$ and 
${\bar N_c}$ representations of the remaining gauge group $U(N_c)$. The field $\phi_2$ is an adjoint of the gauge $U(N_c)$ and a singlet of the global group $U(N_f)$.
On the other hand, $\phi_1$, which decouples at low energies, $E<<m$, is an adjoint of the global $U(N_f)$ and a singlet of the gauge group $U(N_c)$.

There is no vortex solution for $N_f<N_c$. In addition, for $N_f>N_c$, the strings are semi--local[\SEMI,\AUZ] (in the language of $N_c=1$) and may have arbitrary size. This means that they can 
expand without limit and dissolve. (These can lead to stable vortices in the limit $g_2 \to \infty$.)
Therefore, stable non--Abelian strings require $N_f=N_c$. We can realize the case with $N_f=N_c=N$ by wrapping
an equal number of D5 branes on the two nodes. We end up with $N$ flavors of $Q_{12},Q_{21}$ in the $N, {\bar N}$ representations of the $U(N)$ gauge group and an adjoint 
$\phi_2$ which is a singlet of the flavor group. 
The physics is described by the generalization of eqs. (15)-(22) for $U(N)$ with $N_f=N$ flavors which means that they are modified simply by the addition of the trace $Tr$ operator where needed. 
The non--Abelian vortex solutions are obtained by embedding the solution given by eqs. (23)-(25) into any one of the $U(2)$ subgroups of $U(N)$. 
It is easy to see that the non--Abelian vortex has the same tension as the Abelian one, i.e. $T_F=4 \pi ma$ (for vortex number $n=1$). 
Clearly, by the same reasoning above, we can show that the non--Abelian vortex is also a D3 brane wrapped on the second node.

Non--Abelian D--term strings can be obtained by wrapping multiple D3 branes on the nodes of the singularity in eq. (35) and repeating the steps in section 3.2. This is a straightforward
exercise which we leave to the reader. BPS properties of non--Abelian F or D--term strings are the same as those of their Abelian counterparts. Therefore in order to get non--Abelian BPS
F--term strings, we can wrap multiple D3 branes on the untwisted singularity in eq. (32). On the other hand, non--Abelian BPS D--term strings are obtained by wrapping multiple D5 branes on the
singularity in eq. (35).

As an interesting point, we note that the moduli space of a non--Abelian ($U(N)$) vortex is $R^2 \times CP^{N-1}$ where $R^2$ and $CP^{N-1}$ parametrize the location of the vortex in 
two dimensional transverse space and the orientation of the $U(2)$ subgroup in $U(N)$ respectively. The size of $CP^{N-1}$ is $4 \pi/g_2^2$. Using eq. (2) we find that this is equal to 
$V_2/(2 \pi)^2 g_s \ell_s^2$ i.e. the volume of the second node in string units.

\bigskip
\centerline{\bf 4. Domain Walls}
\medskip

Finally, in this section, we describe domain walls[\DW,\TONG,\ONE,\TWO] which are two dimensional solitons in $3+1$ dimensions.
In order to obtain domain walls we need to enlarge the singularity to a deformed $A_3$ singularity fibered over $C(x)$ defined by
$$uv=z(z+m(x-a))(z+m(x-a))(z-mx) \eqno(42)$$
In this case, the gauge group is $U(1)_1 \times U(1)_2 \times U(1)_3$ and there are three singlets $\phi_{1,2,3}$ arising from each of the three nodes. 
We can decouple $U(1)_1$ and $U(1)_3$ from matter
by taking $g_1$ and $g_3$ to be very small. As before we accomplish this by taking $V_1>>\ell_s^2$ and $V_3 >>\ell_s^2$ respectively. In addition, there are two pairs of bifundamentals
$Q_{12},Q_{21},Q_{23},Q_{32}$. Under $U(1)_2$ which is the only remaining gauge group these have charges $1,-1,-1,1$ respectively.
The superpotential is given by
$$W={m \over 2} (\phi_1-a)^2-{m \over 2} (\phi_3-{a \over 2})^2 + Q_{12}Q_{21}(\phi_2-\phi_1) + Q_{23}Q_{32}(\phi_3-\phi_2) \eqno(43)$$
We see that $\phi_1$ and $\phi_3$ are massive and decouple at low energies $E<<m$. (The bifundamentals also get masses of order $a$ but we keep them in the spectrum by assuming $a<E<<m$.)
This is done by setting their F--terms
$$F_{\phi_1}=m(\phi_1-a)-Q_{12}Q_{21} \eqno(44)$$
$$F_{\phi_3}=-m(\phi_3-{a \over 2})+Q_{23}Q_{32} \eqno(45)$$
to zero. Thus we get
$$\phi_1={{Q_{12}Q_{21}} \over m}+a \qquad \phi_3={{Q_{23}Q_{32}} \over m}+{a\over 2} \eqno(46)$$
Substituting the above VEVs into eq. (43) we get the low--energy superpotential
$$W=Q_{12}Q_{21} \left(\phi_2-a-{{Q_{12}Q_{21}} \over {2m}} \right)+Q_{23}Q_{32} \left({a \over 2}+{{Q_{23}Q_{32}} \over {2m}}-\phi_2 \right) \eqno(47)$$
which leads to the F--terms
$$F_{Q_{12}}=Q_{21} \left(\phi_2-a- {{Q_{12}Q_{21}} \over m} \right) \eqno(48)$$
$$F_{Q_{21}}=Q_{12} \left(\phi_2-a-  {{Q_{12}Q_{21}} \over m} \right) \eqno(49)$$
$$F_{Q_{23}}=Q_{32} \left({a \over 2}+ {{Q_{23}Q_{32}} \over m}-\phi_2 \right) \eqno(50)$$
$$F_{Q_{32}}=Q_{23} \left({a \over 2}+ {{Q_{23}Q_{32}} \over m}-\phi_2 \right) \eqno(51)$$
$$F_{\phi_2}=Q_{12}Q_{21}-Q_{23}Q_{32} \eqno(52)$$
In addition, we blow up the second node in order to get an anomalous D--term for $U(1)_2$ where $\xi=\int_{S_2^2}J$. The D--term becomes
$$D=(|Q_{12}|^2-|Q_{21}|^2-|Q_{23}|^2+Q_{32}|^2+\xi) \eqno(53)$$
From the F and D--terms above, we find that the scalar potential has two isolated supersymmetric vacua at 
$$\phi_2=a \qquad |Q_{21}|^2=\xi \qquad Q_{12}=Q_{23}=Q_{32}=0 \eqno(54)$$ 
and 
$$\phi_2={a \over 2} \qquad |Q_{23}|^2=\xi \qquad Q_{12}=Q_{21}=Q_{23}=0 \eqno(55)$$
As a result, there is a domain wall which interpolates between these two vacua. In the world--volume field theory, the domain wall solution (in the limit $a^2>>g_2^2 \xi$) is given by
$$Q_{21}={\sqrt{\xi} \over A} e^{-az} \qquad Q_{23}={\sqrt{\xi} \over A} e^{az/2} \qquad \phi_2={a \over 4}(3-tanhz) \eqno(56)$$
where $A^2=e^{-2az}+e^{az}$ and for simplicity, we assumed that the domain wall is normal to the $z$ direction and located at $z=0$.
The tension of the wall is
$$T_w=\xi(a-{a \over 2})={1 \over 2}\xi a \eqno(57)$$
The brane configuration that corresponds to the domain wall is a D5 brane wrapped on the second node and stretched between $x=a/2$ and $x=a$, i.e. between the two D5 branes wrapped
on the first and third nodes. The tension of such a brane is
$$T=T_{D5} V_2 (a-a/2) (2 \pi) \ell_s^2={V_2 \over {(2 \pi)^5 g_s \ell_s^6}} (\pi a \ell_s^2)= {1 \over 2}\xi a \eqno(58)$$
using eq. (40) for $\xi$. For the domain wall to exist at low energies, $E<<m$, we need to assume $\xi a<<m^3$.

We also note that outside the domain wall, i.e. for either one of the solutions in eqs. (53) or (54), $U(1)_2$ is spontaneously broken by the charged field VEVs. Therefore, we expect to find
vortex solutions outside the domain walls. The two vacua in eqs. (53) and (54) lead to the same vortices with tension $T_s=2 \pi \xi$. These are D--term strings since their tension arises
from an anomalous D--term. Thus, there are vortices on either side of the domain wall.
These strings are semi--local since the gauge group is $U(1)_2$ but there are effectively two flavors. Thus, we expect them to be unstable and dissolve. We can try to stabilize these
strings by not decoupling either $U(1)_1$ or $U(1)_3$. 

For example, if we decouple only $U(1)_3$, the remaining gauge group becomes $U(1)_1 \times U(1)_2$ and in the vacuum given by eq. (54) the gauge symmetry gets enhanced to $U(2)$ which is
spontaneously broken by the VEVs of $Q_{12},Q_{21}$. This leads to non--Abelian vortices with $N_c=N_f=2$ which are stable. In the other vacuum given by eq. (55) which is 
on the other side of the domain wall, vortices are still semi--local and unstable. Clearly, if we do not decouple any of the $U(1)$s, the vortices on both sides of the domain wall 
become stable non--Abelian strings with $N_c=N_f=2$.  

In additon, as in section 2, we expect to find monopoles living inside the vortices. From eq. (54) we see that $U(1)_1 \times U(1)_2$ is broken everywhere except inside the 
D--term strings where $Q_{12}=Q_{21}=0$. Then, $\phi_2$ can have a VEV different than $a$ leading to two types of monopoles charged under $U(1)_1$ or $U(1)_2$ with masses 
$m_m=4 \pi (\phi_2-a)/g_i^2$ where $i=1,2$.  Thus, we expect to find different types of confined monopoles that live in the vortices  which are confining magnetic flux tubes. If we do not decouple 
$U(1)_3$ we get two other types of monopoles charged under either $U(1)_2$ or $U(1)_3$ with masses $m_m=4 \pi (\phi_2-a/2)/g_i^2$ where $i=2,3$.


We see that the physics of domain walls is quite rich even in the simplest case we considered above. Generalizing our results to $A_n$ singularities and/or the non--Abelian case 
with mulptiple D5 branes wrapped on the nodes of the singularity is expected to lead to richer and more interesting results.


\bigskip
\centerline{\bf 5. Conclusions and Discussion}
\medskip

In this paper, we described monopoles, vortices and domain walls that live on the world--volumes of D5 branes wrapped on $A_n$ type singularities.
Monopoles are D3 branes wrapped on the simplest singularity, namely an $A_1$ (which is neither deformed nor fibered) and stretched along $C(x)$. 
Vortices are D3 branes wrapped on deformed and/or resolved $A_2$ singularities fibered over a complex plane $C(x)$. 
They require a singularity with at least two nodes since the charged matter
fields that are needed to spontaneously break the $U(1)$ gauge group arise from strings that stretch between branes wrapped on these two nodes. 
F and D--term strings arise from the deformation and resolution of the fibered $A_2$ singularities respectively.
On the other hand, domain walls are D5 branes wrapped on a deformed singularity with at least three nodes, i.e. a deformed $A_3$ with a resolved node.
Our results can be easily generalized to the cases of multiple D5 branes on deformed $A_N$ singularities such that between any two neighboring
nodes there is a different monopole or vortex and among any three nodes there is a domain wall (with or without strings attached). Such a brane configuration would give rise to a collection of solitons on the noncompact world--volume.

At first sight, our results are somewhat surprising. For example, it is well-known that a D3 brane inside a D5 brane constitutes
a magnetic flux tube[\DOU]. One expects this to hold even when both branes are compactified on a $T^2$ since T--duality relates this to a configuration with a D1 brane inside a D3 brane. 
However, in our case, we compactify both branes on $S^2$ rather than on $T^2$ and T--duality does not apply. The same argument can be repeated for our description of monopoles and domain walls. 
Again, it is well--known that the end of a D3 brane that ends on a D5 brane behaves as a monopole on the D5 brane world--volume. This configuration describes our
monopoles on the singularity even though both branes are wrapped on $S^2$ (rather than on $T^2$ which would allow T--duality). Domain walls on $A_3$ give rise to the same puzzle.
Our results strongly indicate that wrapping branes on nodes of singularities or $S^2$s somehow has the same effect as wrapping them on tori, at least as far as the existence of 
solitons are concerned. It would be interesting to clarify this puzzle by dimensionally reducing world--volume theories of D5 
branes wrapped on $S^2$ and finding out the different soliton solutions. 

The world--volume theories and other quantum properties of vortices and domain walls have been obtained in the framework of intersecting branes[\HAN-\MOD]. 
These describe the low--energy physics of the solitons
in terms of their moduli. For D5 and D3 branes wrapped on the deformed and resolved nodes we expect to find similar world--volume theories. Since our setup
of branes on $A_n$ singularities is related to intersecting brane models this would not be very surprising. For example, the $A_2$ singularity corresponds to three NS5 branes separated
along a compactified $x_6$ direction. (This can be seen as three NS5 branes where the first and third ones are identified.)
The D5 branes wrapped on the two nodes correspond to D4 branes stretched between the two NS5 branes. Singlet masses are obtained by rotating
one of the NS5 branes whereas F and D--terms are obtained by shifting one of the NS5 branes along the $x_8+ix_9$ and $x_7$ directions respectively. Thus, our models can also be described in terms of intersecting branes.

However, the models we examined in this paper are not exactly dual to the intersecting brane models of refs. [\HAN-\MOD].
The main difference is the amount of supersymmetry; the vortices and domain walls (but not the monopoles) above were obtained in theories with ${\cal N}=1$ supersymmetry whereas those in refs. [\HAN-\MOD] arise in ${\cal N}=2$ supersymmetric theories.
Another important difference seems to be the existence of D6 banes in these intersecting brane constructions. The models considered in this paper are dual to intersecting brane models that contain 
semi--infinite D4 branes (which correspond to the D5 branes wrapped on the decoupled nodes) rather than D6 branes. Therefore, we cannot directly use the results in refs. [\HAN-\MOD] especially for describing the soliton moduli spaces or world--volume theories.

It would be worthwhile to find out the world--volume theories for the vortices and domain walls described in this paper. Since, our models have only ${\cal N}=1$ supersymmetry, this will 
involve the collective coordinates of the solitons which are flat directions that are analogous to the moduli in ${\cal N}=2$ supersymmetric theories. These theories should help us understand the low--energy physics of the above solitons and their interactions.

\bigskip

\centerline{\bf Acknowledgements}

I would like to thank the Stanford Institute for Theoretical Physics for hospitality.

\vfill

\refout

\end
\bye